\documentclass[prb,aps,reprint,longbiblography]{revtex4-2}
\usepackage{amsmath, amssymb, graphicx, braket, float, quantikz}
\usepackage{todonotes}
\usepackage{subcaption,graphicx}
\usepackage[colorlinks,allcolors=blue]{hyperref}
\usepackage[noabbrev,capitalize]{cleveref}
\usepackage{xcolor}
\begin{document}

\title{Ground state energy and phase transitions of the long-range XXZ chain using variational quantum eigensolver}

\author{Mrinal Dev}
\email{mrinaldev3@gmail.com}
\affiliation{Department of Physics and Astronomy, National Institute of Technology Rourkela, Rourkela—769008, India}
\author{Shraddha Sharma}
\email{sharmas@nitrkl.ac.in, shrdha1987@gmail.com}
\affiliation{Department of Physics and Astronomy, National Institute of Technology Rourkela, Rourkela—769008, India}
%\email{your-email@example.com}

\begin{abstract}
The variational quantum eigen solver (VQE), has been widely used to find the ground state energy of different Hamiltonians with no analytical solutions and which are classically difficult to compute. In our work, we have for the first time used VQE to identify the phase transition boundary for an infinite- order phase transition. Typically, in a finite-order phase transition, finite-order derivatives of the ground state energy would signal the phase transition. However, this is not the case for the infinite-order phase transitions, where a globally-ranged order parameter is required. In this work, we use a long-range XXZ (LRXXZ) chain for our study. It has been observed that there exist two types of phase transitions for this model. One of which is a first-order phase transition, straightforwardly evaluated by the gradient of the ground energy. Second, an infinite-order phase transition, which is conventionally unfeasible to evaluate using the ground state energy. Therefore, it has been observed that the ground-state energy is not sufficient to probe both of these transition boundaries. However, we propose a simple technique to utilise the ground state energy from VQE to identify both the phase transitions. The idea rests on the argument that VQE requires an ansatz circuit; therefore, the accuracy of the VQE will rely on this ansatz circuit. We have designed the ansatz circuit such that the estimated ground state energy is sensitive to the phase it is evaluated in. It is achieved by applying the constraint that the net spin remains constant throughout the optimization process. Consequently, the ansatz works in a certain phase where it gives relatively small random error, as it should, when compared to the error in ground state energy calculations of the other phases, where the ansatz fails. Identifying these changes in the behaviour of the error in ground state energy evaluation using VQE, we were able to identify the phase boundaries. Using exact diagonalisation, we also compare the behaviour of the energy gradient and energy gap across both the phase transition boundaries for this model. Further, by increasing the depth of the optimisation circuit, we also accurately evaluate the ground energy of the LRXXZ chain for $J>0$ and $J<0$, in the paramagnetic region.
\end{abstract}

%\begin{abstract}
%VQE, the variational quantum eigen solver, has been widely used to find the ground state energy of different Hamiltonians, whose ground state has no analytical solutions and is classically difficult to compute. In our work, we have for the first time used VQE to find phase transition boundaries. We have chosen LRXXZ long-range XXZ for our study. There are two phase transitions for LRXXZ. One is a first-order phase transition that can be easily evaluated by the gradient of the ground energy. Second, an infinite order phase transition is difficult to evaluate. People have used geometric entanglement, tensor networks, and other numerical methods. Analytically, the mean field approximation has been used, as an exact analytical result is not possible. We propose a very simple way to find both the phase transitions. VQE requires an ansatz circuit; the accuracy of the VQE will depend on this ansatz circuit. We have designed the ansatz circuit such that the energy evaluated is sensitive to the phase we are evaluating it in.
%The ansatz works in a certain phase where it gives random error as a good evaluation normally should, but fails definitely in the other phase and gives predictable errors. Identifying these changes in the behaviour of error, we were able to identify the phase boundaries. We also accurately evaluated the ground energy of long-range XXZ, for which it was originally constructed.
%\end{abstract}

\maketitle

\section{Introduction}
Interacting many-electron systems have been at the forefront of condensed matter physics. \cite{leblanc2015solutions}  One attempts to design and understand properties of useful materials\cite{scalapino2012common} and also to understand fundamental science\cite{arovas2022hubbard}.
In this matter, studying long-range interactions has gained a lot of attraction in recent times because of the many emerging setups in atomic, molecular, and optical physics. There have also been several experimental improvements in the cooling, trapping, and controlling of many-body atomic and molecular quantum systems possessing long-range interactions\cite{lahaye2009physics,schneider2012experimental,monroe2014ion,browaeys2016experimental}. Theoretically, they have also been extensively studied in condensed matter physics\cite{bloch2008many,bloch2012quantum,lewenstein2012ultracold}.
Although  short-range systems can be used to derive properties of a corresponding long-range interacting system (in the limiting case), many novel phenomena are exclusive to the long-range interaction itself \cite{spivak2004phases,lahaye2009physics,peter2012anomalous}.
Experimentally, it has been possible to  engineer a {2D-} Ising lattice with power-law decaying interaction between spins\cite{britton2012engineered}. In the study of ultra-cold atoms, it has been observed that for fermionic and bosonic isotopes of Cr, Dy, and Er, a coherent spin-exchange has been induced by a large dipole-dipole interaction decaying like $\frac{1}{r^3}$ \cite{de2013nonequilibrium,naylor2015chromium,lu2011strongly,baier2016extended,aikawa2014reaching,baumann2010dicke}.

In our work, we have evaluated the ground-state energy of the Long Range XXZ model (LRXXZ) with good accuracy and evaluated its phase transitions using quantum algorithms. Generally, phase transitions are studied using singularities or discontinuities in expectation value of local variables by varying parameters along the critical point or using the two-point correlation function\cite{sachdev1999quantum}. These methods are sufficient for finite-order phase transitions, however, such techniques fail when we have an infinite-order phase transition or a topological transition. These transitions are only accessible through global parameters\cite{DeChiara2018} and can usually  only be accurately obtained  by quantum Monte Carlo or density-matrix
renormalization group methods\cite{Ceperley1995,Schollwoeck2011}.
Entanglement properties can also be very useful in this domain,  von Neumann entropy \cite{Vidal2003,Hastings2007,Calabrese2004,Holzhey1994,Korepin2004} and geometric entanglement \cite{Barnum2001,Shimony1995,Wei2003,Wei2005,Orus2008} have been shown to be good  candidates for detecting these quantum phase transitions.
This domain of phase transitions is yet to be explored by quantum computing. In our work, we propose to use Variational Quantum Eigensolver (VQE) to probe different phase transitions of LRXXZ system. VQE is a very powerful algorithm for condensed matter physics and has been used to find the ground state energy of {complex systems} like the Fermi-Hubbard model \cite{stanisic2022observing}.
{The VQE is conventionally designed to extract the ground state energy; using this technique, we also obtain the ground state energy of the LRXXZ chain with good accuracy}.

\section{Long-Range XXZ Model}
In this section, we will start with the introduction of the model under consideration, i.e., the LRXXZ chain, given by the Hamiltonian,

\begin{equation}
H = -J \sum_{i \neq j}^{N} \frac{S_i^x S_j^x + S_i^y S_j^y + \Delta S_i^z S_j^z}{|i - j|^\alpha},
\end{equation}

\noindent where $J$ is the coupling constant, $\Delta$ is the anisotropy term, whereas, $S^{x,y,z}$ are Pauli spin-$\frac{1}{2}$ operators. The $i,j$ represents site indices, with $i \in \mathbb{Z}$ ($i\in[0,N]$, $N$ being the system size), and the range of interaction is defined by $\alpha$, with $\alpha \in \mathbb{R}$. When $\alpha$ is small, the system is long-range interacting, and as $\alpha$ tends to infinity, it transforms to the short-range XXZ model, implying only nearest-neighbor interaction. The one-dimensional short-range XXZ has been extensively studied \cite{franchini2017introduction} and its phases are well known. It was observed that for $J>0$, the paramagnetic (PM) phase is obtained for $-1<\Delta< 1$, whereas for $\Delta > 1$, the system is in the ferromagnetic (FM) phase, and for $\Delta < -1$ it is in the anti-ferromagnetic (AFM) \cite{franchini2017introduction} phase. However, as the range of interaction is increased, it has been observed that the transition point of AFM to PM phase starts to shift towards more negative $\Delta$ values. In contrast, the PM-FM transition point remains the same\cite{frerot2017entanglement, schneider2022entanglement}. It is to be noted that these classifications of AFM, PM and FM are defined with respect to the ordering of net magnetization in the $z$-direction.

% Interestingly, as the sign of $J$ is changed, i.e., for $J<0$, the phases are flipped for the short-range XXZ Hamiltonian \cite{franchini2017introduction}.%}, and as the range of interaction is increased., these transition lines also display a unique behaviour, which, to our knowledge, has never been explored before.  

\section{VQE and the Ansatz Circuit}
The VQE is a hybrid quantum-classical method that can be used to find the ground-state energy of a Hamiltonian. The expectation value of the Hamiltonian is computed over a parameterized state using a quantum processor, while a classical optimizer adjusts the parameters to minimize energy \cite{cerezo2022variational}. An ansatz state is chosen for parameterization. We start initially with the Neel state, which is prepared by applying the Pauli-$X$ gate alternatingly to the initial default $|000\cdots\rangle$ state. The alternatively applied Pauli-$X$ gates will flip $|0\rangle\to|1\rangle$, leaving us with the $|0101\cdots\rangle$ Neel state as shown in Fig.~(\ref{initialization}). Once the $|0101\cdots\rangle$ is obtained, we apply parameterized gates, {as shown in Fig.~(\ref{fig:variational_circuit}),} leading to our ansatz state. Afterwards, we tune the parameters of these gates to minimize energy, which leads the ansatz state closer and closer to the actual ground state of the system. In simple terms, through the circuit shown in Fig.~(\ref{fig:variational_circuit}), we make a guess at a ground state i.e., the parameterized state. Thereafter, we vary the parameters until the ground state is reached by minimizing the energy. We can construct our ansatz circuit such that the total spin remains constant throughout. This approach, combined with the choice of the initial state, will play a crucial role in finding phase transitions, as we will see later in the paper.

It is therefore clear that this approach heavily relies on the initial guess for our ground state or the ansatz state. The better our guess is, the easier we can reach the ground state. There are several good guesses; one of them is the HVA (Hamiltonian variational ansatz). In HVA, one constructs the ansatz using the Hamiltonian of the system; every term is written in the form of Pauli matrices, and the matrix representation of these terms can be found. Then, these matrices can be implemented using the gates one wishes to use, depending on the hardware of the system. These gates are parameterized, and all terms that commute share parameters as they have the same eigenstate.
For the short-range XXZ Hamiltonian, the ansatz is constructed by separating it into odd and even parts; all the odd parts and even parts are disjoint from each other, so they commute with each other and can be run in parallel with the same parameter \cite{wiersema2020exploring}.
The Hamiltonian can be separated in the following way
\begin{eqnarray}
H^{\text{even}} &=& \sum_{s}H_{ss}^{even}=  \sum_{s}\sum_{i=1}^{N/2} \sigma^s_{2i-1}\sigma^s_{2i},\nonumber \\
H^{\text{odd}} &=& \sum_{s} H_{ss}^{odd}= \sum_{s}\sum_{i=1}^{N/2-1} \sigma^s_{2i}\sigma^s_{2i+1},
\end{eqnarray}

\noindent
where $s \in \{x,y,z\}$ denotes the spin component. Each of the spins has different parameters, but each even bond and each odd bond will have the same parameter because they are disjoint
For $N=4$,\\
\noindent
Even bonds: $(1,2), (3,4) \;\;\rightarrow\;\; \sum_{s\in{ x,y,z}}(\sigma^{s}_1\sigma^s_2 + \sigma^{s}_3\sigma^s_4)$.\\
\noindent
Odd bonds: $(2,3) \;\;\rightarrow\;\;  \sum_{s\in{ x,y,z}}(\sigma^{s}_2\sigma^s_3)$.\\
\medskip

%\noindent
%So we define
%\[
%H^{\text{even}} = h_{12} + h_{34}, \qquad 
%H^{\text{odd}} = h_{23}.
%\]
We can put all the terms in a single equation representing the implementation of the ansatz circuit as:

\begin{eqnarray}
U_{XXZ}(\boldsymbol{\beta}, \boldsymbol{\gamma}, \boldsymbol{\theta}) 
&=& \prod_{\ell=1}^{p}
G(\theta_\ell, H_{zz}^{\text{odd}})
G(\theta_\ell, H_{zz}^{\text{even}}) \nonumber \\
&~& \times G(\gamma_\ell, H_{xx}^{\text{odd}})
G(\gamma_\ell, H_{xx}^{\text{even}}) \nonumber \\
&~& \times G(\beta_\ell, H_{yy}^{\text{odd}})
G(\beta_\ell, H_{yy}^{\text{even}}),\\
&~& \nonumber\\
\text{with,}~G(\theta, H) &=& e^{-i \theta H/2}
\end{eqnarray}

% \begin{equation}
% G(\theta, H) = e^{-i \theta H}.
% \end{equation}

Like HVA, we have also used the Hamiltonian for our inspiration to construct the circuit. In the variational ansatz, the Hamiltonian is decomposed into even and odd parts, and these parts are parameterized separately. 
We have used a different approach to construct the ansatz. Let us first look at our Hamiltonian:
\begin{equation}
e^{-i H t} \rightarrow e^{-i t (S_i^x S_{i+1}^x + S_i^y S_{i+1}^y + \Delta S_i^z S_{i+1}^z)}
\end{equation}
Since $[S_i^x S_{i+1}^x + S_i^y S_{i+1}^y, S_i^z S_{i+1}^z] = 0$, we can write:
\begin{equation}
e^{-i t (S_i^x S_{i+1}^x + S_i^y S_{i+1}^y + \Delta S_i^z S_{i+1}^z)} = e^{-i \theta (S_i^x S_{i+1}^x + S_i^y S_{i+1}^y)} e^{-i \phi S_i^z S_{i+1}^z}
\end{equation}
We parameterize time with $\theta$ and $\phi$.
Since we know the matrix representation of $S^{s \in \{x,y,z\}}$, we can use them to find the matrix representation of each part as:
\begin{equation}
e^{-i \theta (S_i^x S_{i+1}^x + S_i^y S_{i+1}^y)} e^{-i \phi S_i^z S_{i+1}^z}
=
\begin{array}{c}
\begin{bmatrix}
1 & 0 & 0 & 0 \\
0 & \cos\theta & -i \sin\theta & 0 \\
0 & -i \sin\theta & \cos\theta & 0 \\
0 & 0 & 0 & 1
\end{bmatrix} \\[6pt]
\times
\begin{bmatrix}
e^{-i\phi} & 0 & 0 & 0 \\
0 & e^{i\phi} & 0 & 0 \\
0 & 0 & e^{i\phi} & 0 \\
0 & 0 & 0 & e^{-i\phi}
\end{bmatrix}.
\end{array}
\label{matrixrepersentation}
\end{equation}
We further show in Eqs.~(\ref{circuit1}) and (\ref{cicruit2}) the circuit implementation for these matrices.

\begin{equation}
\begin{array}{ccc}
\begin{bmatrix}
1 & 0 & 0 & 0 \\
0 & \cos\theta & -i \sin\theta & 0 \\
0 & -i \sin\theta & \cos\theta & 0 \\
0 & 0 & 0 & 1
\end{bmatrix}
& \rightarrow &
\begin{quantikz}
\lstick{$q_0$} &  \ctrl{1}   & \gate{R_x(\theta)} & \ctrl{1} & \qw \\
\lstick{$q_1$} &\targ{} & \ctrl{-1} & \targ{} & \qw
\end{quantikz}
\end{array}
\label{circuit1}
\end{equation}
\begin{equation}
\begin{array}{ccc}
\begin{bmatrix}
e^{-i\phi} & 0 & 0 & 0 \\
0 & e^{i\phi} & 0 & 0 \\
0 & 0 & e^{i\phi} & 0 \\
0 & 0 & 0 & e^{-i\phi}
\end{bmatrix}
& \rightarrow &
\begin{quantikz}
\lstick{$q_0$} & \ctrl{1}      & \gate{R_z(\theta)}  &  \ctrl{1}      & \qw \\
\lstick{$q_1$} & \targ{}& \qw                       & \targ{}& \qw
\end{quantikz}
\end{array}
\label{cicruit2}
\end{equation}
Afterwards, these gates are applied to all the qubits in the system, connecting each of the neighbouring qubits, which makes just 1 layer of the circuit (depth=1). The ansatz circuit is repeated up to the desired depth (=2, in our case). However, after implementation, we found that the ansatz circuit did not achieve the ground state very effectively, so we modified the RZ gate with the CRZ gate. These modifications are allowed because the ansatz circuit is just an ansatz circuit for finding the ground state energy and need not follow any particular rule, and the guess that works the best may be chosen. The final ansatz circuit after the application of an alternate $X$-gate is represented in Fig.~(\ref{fig:variational_circuit}).
\begin{figure}[h]
\centering
\begin{quantikz}
\lstick{$q_0: \ket{\uparrow}$} & \qw & \qw & \qw & \rstick{$\ket{\uparrow}$ } \\
\lstick{$q_1: \ket{\uparrow}$} & \gate{X} & \qw & \qw & \rstick{$\ket{\downarrow}$ } \\
\lstick{$q_2: \ket{\uparrow}$} & \qw & \qw & \qw & \rstick{$\ket{\uparrow}$}
\end{quantikz}
\caption{Circuit representation of initialisation using three-qubits with alternate application of the X-gate.}
\label{initialization}
\end{figure}

\begin{figure*}[t]
\centering
\begin{quantikz}
\lstick{$|0\rangle_{0}$} & \ctrl{1} & \gate{RZ(z_{0,0})} & \ctrl{1} & \qw & \qw & \qw & \ctrl{1} & \gate{RX(-xy_{0,0})} & \ctrl{1} & \qw & \qw & \qw \\
\lstick{$|1\rangle_{1}$} & \targ{} & \ctrl{-1} & \targ{} & \ctrl{1} & \gate{RZ(z_{0,1})} & \ctrl{1} & \targ{} & \ctrl{-1} & \targ{} & \ctrl{1} & \gate{RX(-xy_{0,1})} & \ctrl{1} \\
\lstick{$|0\rangle_{2}$} & \qw & \qw & \qw & \targ{} & \ctrl{-1} & \targ{} & \qw & \qw & \qw & \targ{} & \ctrl{-1} & \targ{}
\end{quantikz}
\caption{Three-qubit ansatz circuit for a single layer.}
\label{fig:variational_circuit}
\end{figure*}

We have parameterized every single gate with a different parameter. This is unlike conventional variational ansatz, where commuting terms share the same parameters; however, doing this allows us to achieve the expected ground state energies with very low circuit depth of $p = 2$, compared to previous studies \cite{wiersema2020exploring}. Through this method, we find the number of parameters required in total is $4(N-1)$, \textit{i.e.} $2(N-1)$ parameters per layer. Conventionally, an even-odd variational ansatz has $2(N-1)$ parameters , for example, the XXZ model requires 4 parameters per layer with $(N-1)/2$ total layers. However, our circuit has a much lower circuit depth, so the number of gates required is also much less comparatively. 
The number of gates is equal to the number of parameters, but for HVA, the number of gates can be multiple times the number of parameters.
The main advantage is that it works efficiently with LRXXZ as well.

We have not only used this method to find the ground state energy of LRXXZ efficiently for $J<0$, but also, for the first time, used a quantum algorithm to explore the long-range phase transitions for the case of $J>0$. In order to do this, we exploited our initial state coupled with the fact that the ansatz circuit is designed in a way that it does not cause the net magnetization to change. We evaluated the error between the ground state energy obtained via exact diagonalization and the energy evaluated by VQE.  For VQE,  we have used the COBYLA optimiser with 4000 iterations. We started with random initial parameters, and along with the large number of iterations, we can avoid local minima efficiently and ensure good quality convergence. Each data file was then averaged over five runs for both AFM-PM and PM-FM phase boundaries. The error evaluated from this method showed dependency on the phase in which it was evaluated.  We exploited this feature to mark the phase transitions. We will take a closer look at these methods in the results section. 

\section{Results}
In this section, we provide the details of our novel scheme to probe different phase transitions in the LRXXZ model for $N=12$, considering open-boundary conditions. For a system size of $N=12$, the VQE and ED provide a reasonably good approximation to the ground state of the XXZ model \cite{wang2023scalable,de2018study}. Furthermore, a system size dependence has also been performed for checking the robustness of the method in the Appendix (\ref{size}).
 In Sec.~(\ref{PT_LRXXZ}), utilising directional coherence, we probe the phase transition boundaries between FM-PM and PM-AFM phases and compare the two phase transitions for depth$=1$, and  $J=1$. Furthermore, in Sec.~(\ref{dep2}) we also present our results of VQE for depth$=2$, $J=\pm1$, and plot the relative error for the whole phase diagram to highlight the accuracy of the VQE approach utilised in this work.

We start our discussion by introducing the idea of directional coherence as a useful probe for detecting phase transitions.
\subsection{Phase Transition of Long-Range XXZ chain}\label{PT_LRXXZ}

We start by analysing the energy difference of the exact diagonalisation results and the ground energy computed through VQE ($E_d (\Delta,\alpha)=E_{exact}-E_{VQE}$) at depth 1  as a function of $\Delta$ and $\alpha$. The depth is chosen to be 1 because, in this section, our goal was to obtain phase transitions rather than to achieve ground energy with accuracy. We further computed the gradient of this energy difference along $\alpha$ and $\Delta$, and utilized it to compute normalized gradient vectors represented as arrows in our phase diagrams. We divide the $\alpha$ and $\Delta$ values in a $20\times20$ grid and use NumPy’s np.gradient for the calculation of these gradients. The components of these gradient vectors for each $\Delta$ and $\alpha$ can be calculated as:

\begin{eqnarray}
    u_x&=&\frac{\partial{E_d}/\partial{\alpha}}{\sqrt{(\partial{E_d}/\partial{\alpha})^2+(\partial{E_d}/\partial{\Delta})^2}}\nonumber\\
     u_y&=&\frac{\partial{E_d}/\partial{\Delta}}{\sqrt{(\partial{E_d}/\partial{\alpha})^2+(\partial{E_d}/\partial{\Delta})^2}}.\label{uxuy}
\end{eqnarray}

\noindent The main objective in calculating the gradient vector is to finally obtain the information of \textit directional coherence. For this purpose, we obtain the angle of the gradient vector, as:

\begin{eqnarray}
\theta=\tan^{-1}(u_y/u_x).
\end{eqnarray}

\noindent The directional coherence, then, has been calculated as:
%by taking the mean over the sine and cosines of the angle of the normalised gradient vector, as shown below: %in Eq.~\ref{dc}, with the $\Delta$-axis over a moving window. 

\begin{eqnarray}
    \text{Directional coherence} &=&
\sqrt{\left(\overline{\sin\theta}\right)^2 +
      \left(\overline{\cos\theta}\right)^2}
\label{dc}
\end{eqnarray}

% \end{equation}
\noindent where, $\overline{\sin\theta}$, $\overline{\cos\theta}$, are the means of  $\sin\theta$, $\cos\theta$, over a moving window. The definition used in Eq.~(\ref{uxuy}) does not consider the magnitude of the gradient vector, but rather the normalized components of the gradient. Since we aim to analyze the variation or change in the orientation pattern of the gradient vectors. However, we have also checked our results considering magnitude weighted directional coherence in detail in Appendix (\ref{dir_coh}).
The size of the moving window is taken to be $10\times1$ in all the directional coherence plots to get a smooth and accurate result. Although the results are not strongly sensitive to the precise choice of the window size, a practical guideline is to select a window large enough to reduce local fluctuations in the vector field without significantly altering the underlying structure of the data. In the present case, the parameter space is discretised on a $20\times20$ grid in $(\Delta,\alpha)$, the critical region extends over roughly half of the parameter range; therefore, we choose the averaging window to correspond to half of the grid size along the averaging direction, i.e., $10\times1$. One of the guiding lines for choosing window size is by visualizing at which value the chaotic gradient arrows are separated from the aligned one. Further information and details on the  sensitivity to window size in  presented in Appendix (\ref{window})
%\begin{equation}
%\text{where } 
%\tilde{\sin}\theta, \ \tilde{\cos}\theta
%\text{ are means of } 
%\sin\theta, \cos\theta
%\text{ over a moving window.}
%\end{equation}

 The idea is that the ansatz state works particularly well in one phase, leading to a random lower value of $E_d$, and therefore, $E_d$ is expected to be lower in one phase as compared to that in the other phase. The directional coherence, therefore, helps differentiate between the regions where the nature of the $E_d$ changes is owing to the underlying phase transition. Graphically, directional coherence represents the behaviour of arrows in a particular region. We exploit the pattern in the orientation of these arrows. As an example, we observe that these arrows representing gradient vectors point randomly in one phase (AFM), whereas they exhibit a systematic pattern in another phase (PM/FM). The directional coherence can capture this change in the behaviour of arrows, not only between a random or systematic orientation but also between two systematic orientations in two different directions. This is evident in our phase diagrams where we have overlaid the arrows on the pseudo-colour plots of the directional coherence (Fig.~ ~(\ref{energy_diff_ferro}) and (\ref{dir_afm})). We observed that the direction of these arrows can be understood and is in compliance to a good extent with the existing mean field results \cite{frerot2017entanglement}. For the subsequent subsections, we will consider $J=1$.

\subsubsection{FM to PM phase}
\begin{figure*}
\centering
    \subfloat[\label{energy_diff_ferro}]{%
              \includegraphics[width=0.35\linewidth]{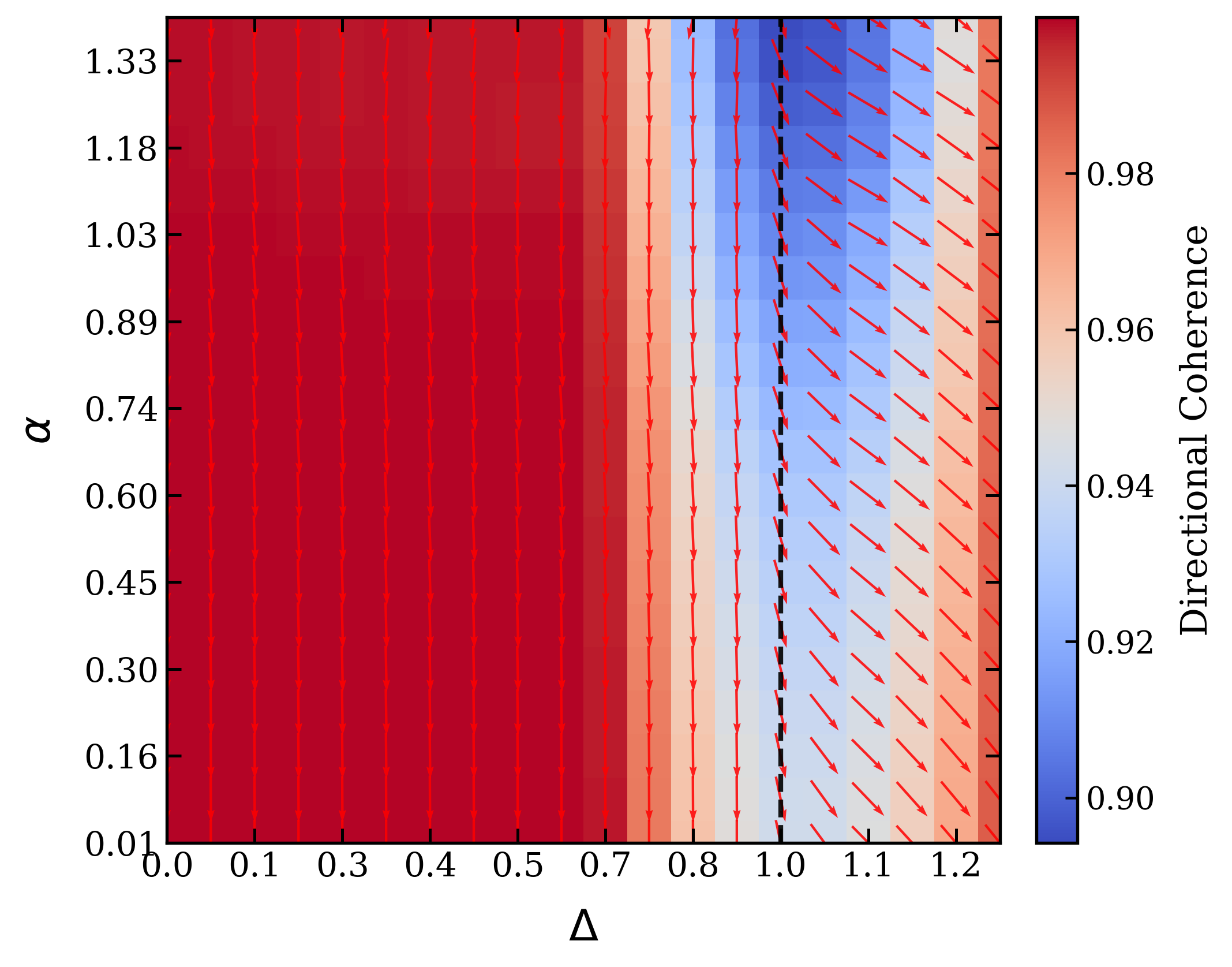}}
    \subfloat[\label{gradient_fm}]{%
               \includegraphics[width=0.35\linewidth]{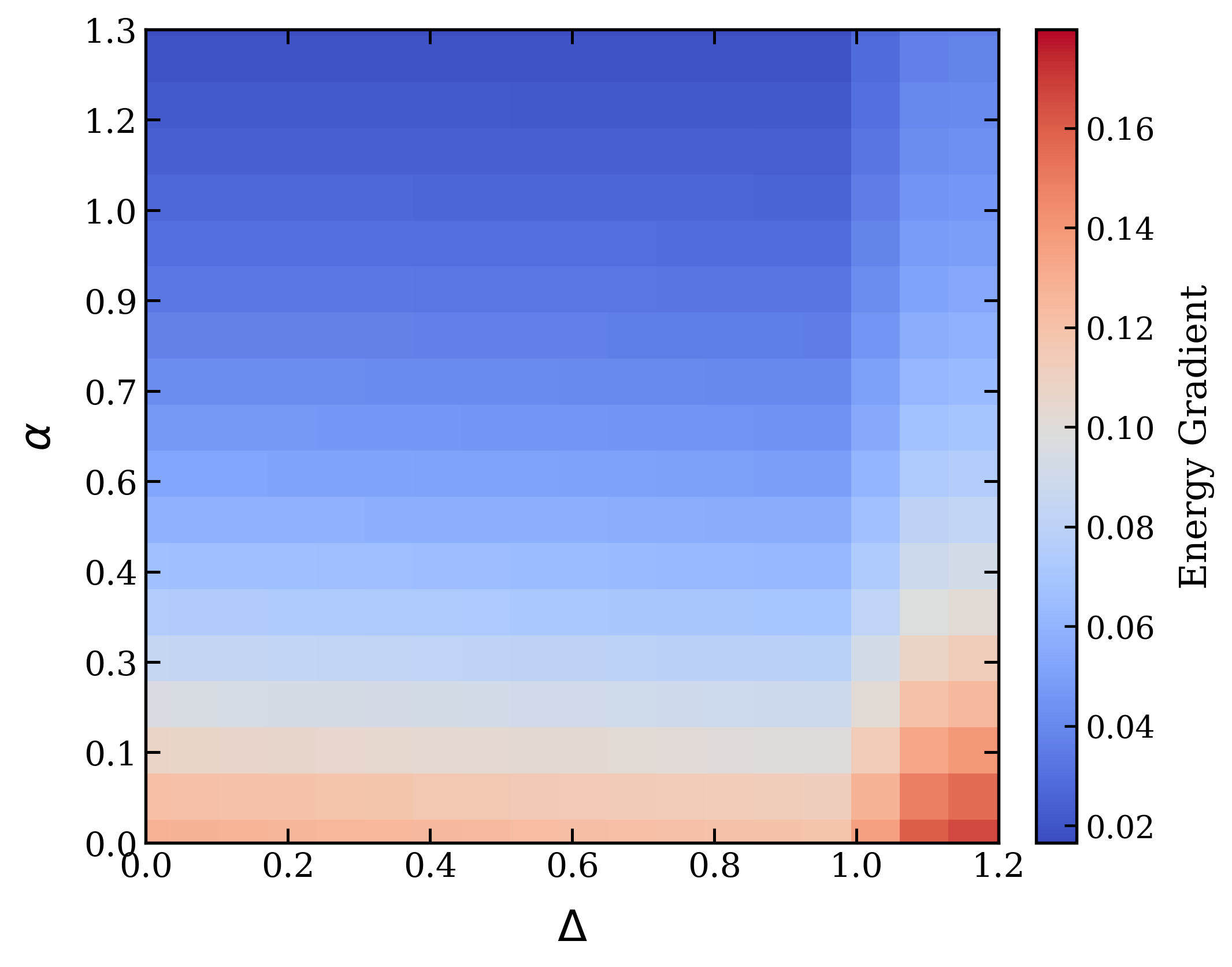}}
         \subfloat[\label{energygap_fm}]{%
               \includegraphics[width=0.35\linewidth]{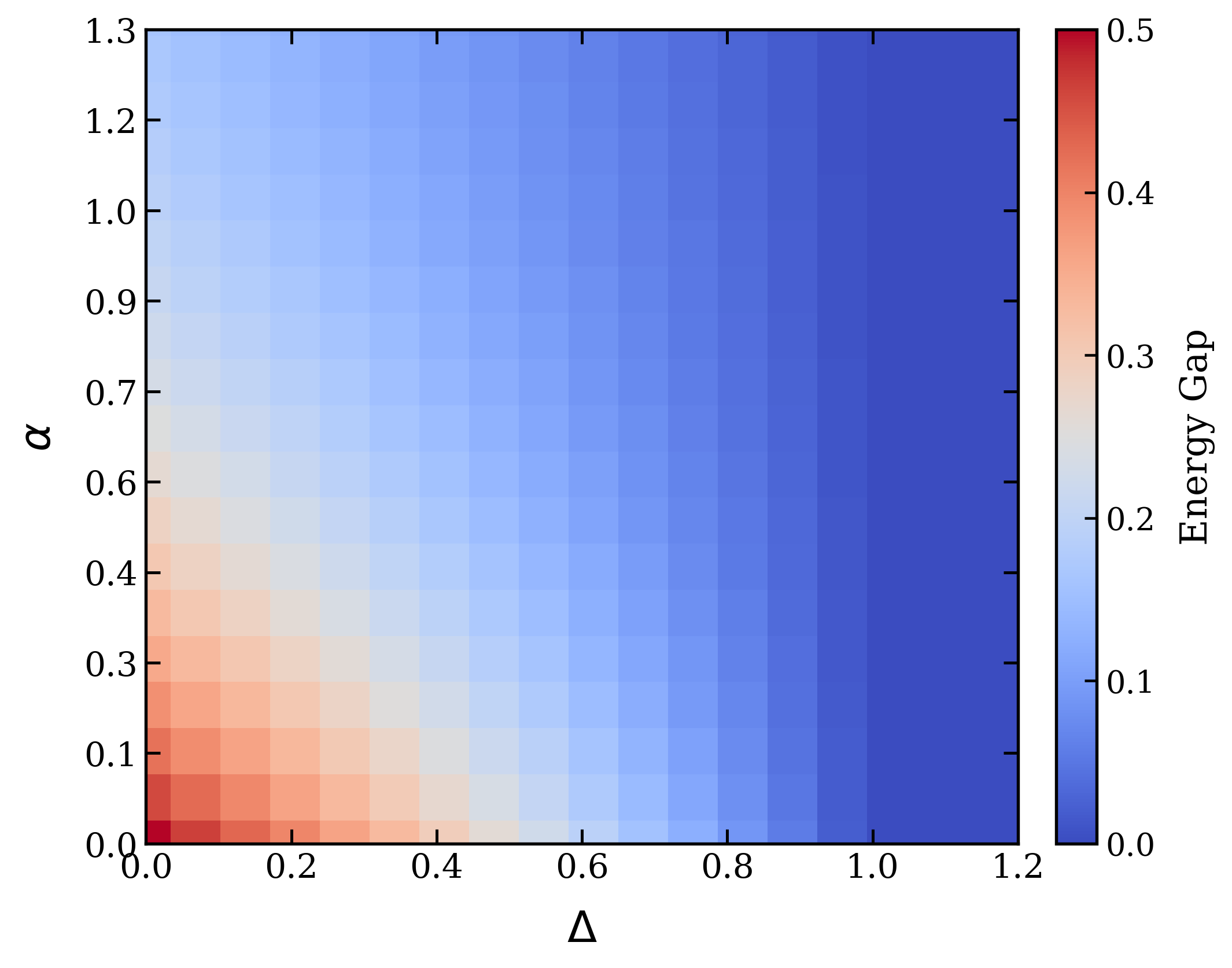}}
    \caption{Phase diagrams illustrating (a) directional coherence, (b) gradient of the ground-state energy, and (c) energy gap as functions of $\Delta$ and $\alpha$. In panel (a), a dotted contour line is shown, highlighting the boundary where a qualitative change in the vector field direction occurs. This contour coincides with $\Delta=1$ marking the phase transition from the PM to the FM phase.}\label{fig:FM_PM}   
\end{figure*}

\begin{figure*}
\centering
    \subfloat[\label{dir_afm}]{%
              \includegraphics[width=0.35\linewidth]{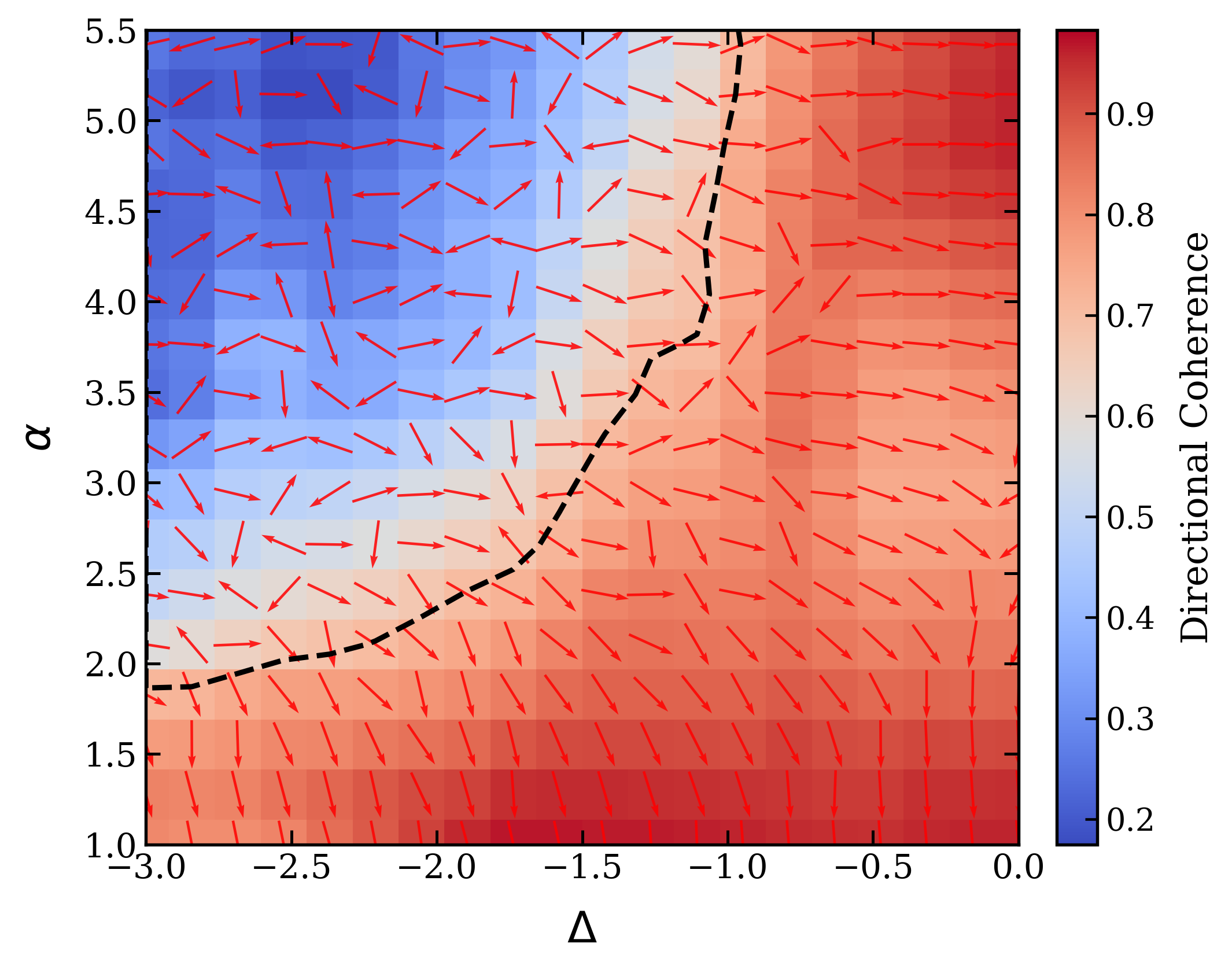}}
    \subfloat[\label{gradient_afm}]{%
               \includegraphics[width=0.35\linewidth]{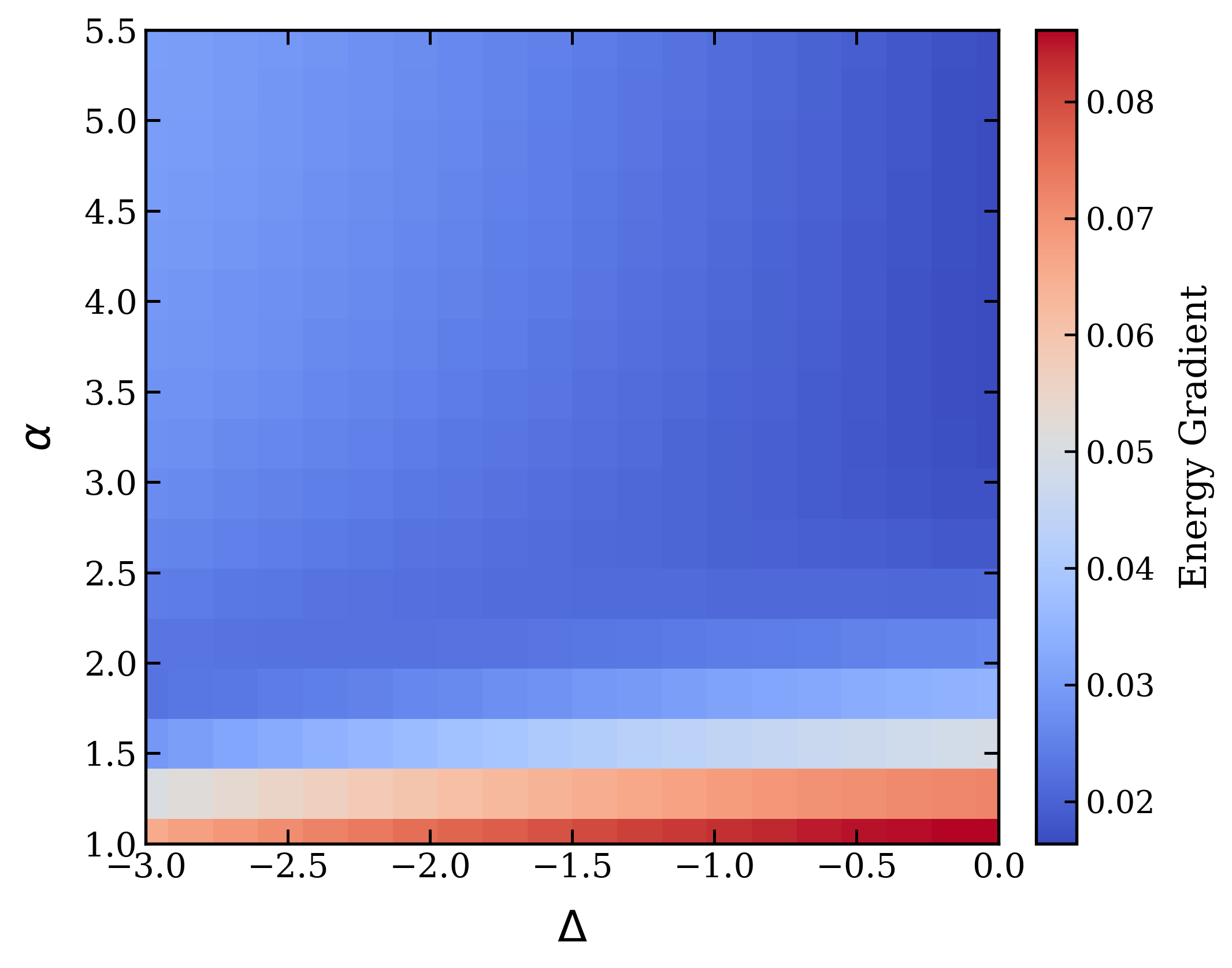}}
         \subfloat[\label{energy_gap_afm}]{%
               \includegraphics[width=0.35\linewidth]{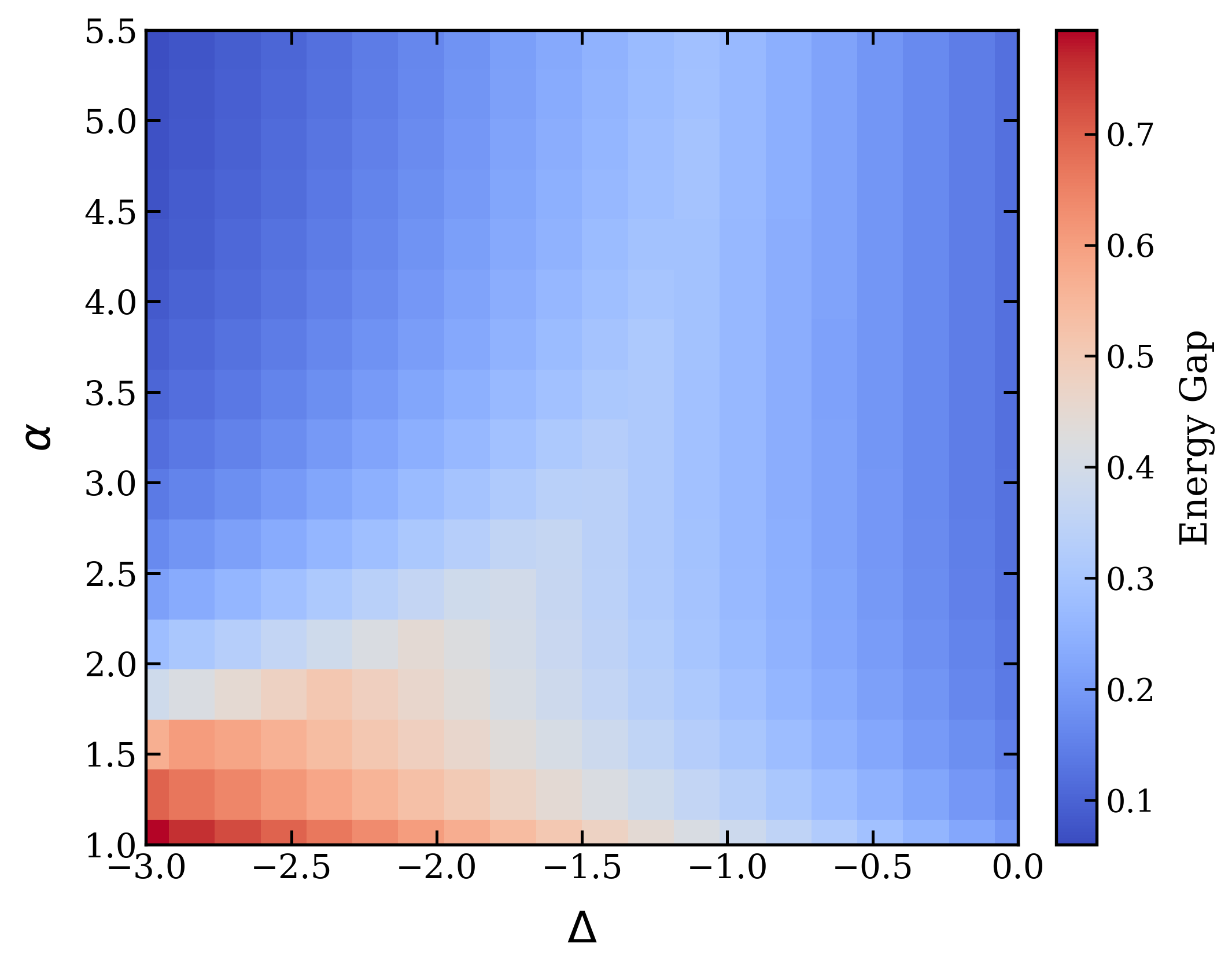}}
    \caption{Phase diagram of (a) directional coherence, (b) energy gradient, and (c) energy gap has been made as a function of $\Delta$ and $\alpha$. Panel (a) shows a clear depiction of AFM-PM phase transition as compared to panels (b) and (c). The dashed line in panel (a) separates the region of randomly oriented vectors (AFM phase, shown in blue) from the coherent region (PM phase, shown in red).}\label{fig:AFM_PM}   
\end{figure*}

For comparison, we explore the behaviour of the energy gap (between the ground and the first-excited state) and the gradient of the ground state energy. These are excellent indicators of any phase transition. The gradient has been computed by finding the exact ground state energy using the exact diagonalisation technique and taking the derivative along $\Delta$ and $\alpha$ as shown in Fig.~(\ref{gradient_fm}). 
The transition from the PM phase to the FM phase\cite{schneider2022entanglement} at $\Delta = 1$ is shown in Fig.~(\ref{gradient_fm}). Left of $\Delta=1$ is the PM phase, and right of it is the FM phase. This transition at $\Delta=1$ can be observed in Fig.~(\ref{gradient_fm}), depicting the gradient of the ground state energy, and in Fig.~(\ref{energygap_fm}) representing the energy gap. It can also be observed from Fig.~(\ref{energy_diff_ferro}) that this PM-FM transition can also be probed using the directional coherence. In Fig.~(\ref{energy_diff_ferro}), we present the behaviour of directional coherence over the whole phase diagram overlaid by gradient vectors. It was observed that the direction of arrows changes abruptly from the PM phase ($\Delta<1$) to the FM phase ($\Delta>1$). Consequently, we see a strip in this region in the directional coherence phase diagram, indicating a phase transition.  One can also comment crudely on the direction of arrows using mean-field theory \cite{frerot2017entanglement} in the FM phase. The energy difference, $E_d$ in this case can be represented as

%%%%%%%%%%%%%%%%%%%%%%%%%%%

\begin{eqnarray}
E_d &= &
\left| (E_{\mathrm{FM}})_{\mathrm{exact}}
- (E_{\mathrm{AFM}})_{\mathrm{approx}} \right| \nonumber\\
&\approx&
\left| (E_{\mathrm{FM}})_{\mathrm{MF}}
- (E_{\mathrm{AFM}})_{\mathrm{MF}} \right|,\label{e_diff}
\end{eqnarray}

\noindent where, $E_{\mathrm{FM}}$ and $E_{\mathrm{AFM}}$ are energies for the FM and AFM phases, respectively. The subscript \textit{approx} is added to reflect that for depth$=1$, the actual ground state energy in AFM phase can be slightly different from the one obtained from VQE. However, from the mean-field point of view, we can approximate $E_d$ as the difference between the energies in the FM and AFM phases. The reason $E_d$ is taken to be the difference between the $E_{\mathrm{FM}}$ and $E_{\mathrm{AFM}}$ is because the ansatz is chosen such that the energy obtained from VQE will be very close to that of the AFM phase (for depth$=1$), whereas the exact diagonalization would lead to $E_{\mathrm{FM}}$. Proceeding in the same line of arguments as in Ref.~[\onlinecite{frerot2017entanglement}], $E_d$ for the case of FM phase will be:

\begin{equation}
\begin{split}
E_d = \frac{|\Delta|}{4} \Bigg| &
\sum_{\substack{i,j \\ i \neq j}}
\left(
\frac{1}{|i-j|^{\alpha}}
\right)
-
\frac{ \epsilon_i \epsilon_j}{|i-j|^{\alpha}}
\Bigg|,
\end{split}
\end{equation}
\noindent where, $\epsilon_i=1 (-1)$ for $i\in \text{even (odd)}$. Therefore, as $\alpha$ is increased, the overall $E_d$ will decrease, but as $\Delta$ is a multiplying factor, the arrows will deviate in the increasing direction of $\Delta$. To summarize, we started with an AFM state in the $z$-direction and used a circuit that kept the net magnetization zero in the $z$-direction. Therefore, our algorithm will never reach FM ground energy and will miserably fail because it will always stay in the zero magnetization state and, which leads to an increase in the error abruptly. The error will increase as $\Delta$ increases, and the arrows will start to point toward the positive $\Delta$ direction. This is a first-order transition, and its clear identification using our technique of directional coherence is a direct consequence of the initial state we started with. One can also start with FM state, in that case there is an accurate evaluation of error in the FM state and high error in the PM state. Further analysis is shown in the Appendix (\ref{ferromagnetic}).
%One can again utilise the mean-field energy calculation arguments to justify this. In the case of PM phase, according to the mean-field approach, similar to before we now have,
% \begin{eqnarray}
% E_d&\approx&\left| (E_{\mathrm{PM}})_{\mathrm{MF}}
% - (E_{\mathrm{AFM}})_{\mathrm{MF}} \right|\nonumber\\
% E_d &=& \frac{1}{4} \Bigg|
% \sum_{\substack{i,j \\ i \neq j}}
% \left(
% \frac{1}{|i-j|^{\alpha}}
% \right)-
% \left(
% \frac{\Delta \epsilon_i \epsilon_j}{|i-j|^{\alpha}}
% \right)
% \Bigg|.
% \end{eqnarray}

% \noindent Therefore, the change in $E_d$ is independent of $\Delta$ and only show the direction aligned w.r.t. $\alpha$. 

% \textcolor{red}{On the other hand, the paramagnetic phase is ferromagnetic in a direction in the paramagnetic plane and allowed to have zero magnetisation in the z-direction \cite{frerot2017entanglement}}

\subsubsection{Paramagnetic phase to AFM phase}
Now we will be looking at a more complicated phase transition of infinite-order. Fig.(\ref{fig:AFM_PM}) shows a phase transition from the PM phase in the right to the AFM phase in the top left corner. This phase transition is also visible in the pseudo-colour plot of the directional coherence and is a consequence of our initial ansatz state, which is an AFM state. Now, in the PM phase, the net magnetisation is still zero, but the AFM order is absent and is more disordered. Our optimizer can still get to the ground state, but since we started with an ordered AFM phase, it is easier for the optimizer to find the ground state in the AFM region than in PM region. However, it still reaches some energy minima, which, as compared to the case of FM phase, is impossible with the same initial ansatz state. Therefore, the increase in error is maximum and abrupt when transitioning to FM from PM phase. However, this change in $E_d$ is gradual during the transition from the AFM to the PM phase. Since this change in error is not very suitable in AFM-PM phase transition case, we evaluated the gradient of the energy difference at each point and found it to be randomly aligned in the AFM region and aligned in some particular direction in the PM region. Consequently, we found the directional coherence of the gradient that separated the region with a random direction gradient from the region where the gradients were aligned in a particular direction.
This allowed us to obtain a clear phase boundary between the two regions, Fig.~(\ref{dir_afm}). We have used dotted lines to separate the randomly aligned region and the coherent region. Through visually looking at directional coherence, we chose the value of 0.7 as the threshold using the colour bar, as that seemed to be the transition point from the red region(coherent) and blue region (random).
This aligns perfectly with previously obtained phase transitions with geometric entanglement \cite{schneider2022entanglement}. The Ref.~[\onlinecite{schneider2022entanglement}] has used an inverse $\alpha$ value, so the transition at $\Delta =-3$ is at inverse $1/\alpha\approx0.56$, which corresponds to $\alpha\approx1.78$. Using our approach, we find the critical $\alpha$ value at $\Delta=-3$ to be $\alpha_c\approx1.75$ with an error margin of 0.06.
However, as shown in Fig.~(\ref {gradient_afm}), this phase transition is not at all identifiable in the ground-state energy gradient phase diagram since it is an infinite-order phase transition. Although it is possible to observe it faintly in the energy gap phase diagram in Fig.~(\ref{energy_gap_afm}), it is not very accurate, and even for that, one needs to find the excited state energy, unlike our approach that probes the ground state energies.

\begin{figure}[ht]
    \includegraphics[width=0.8\linewidth]{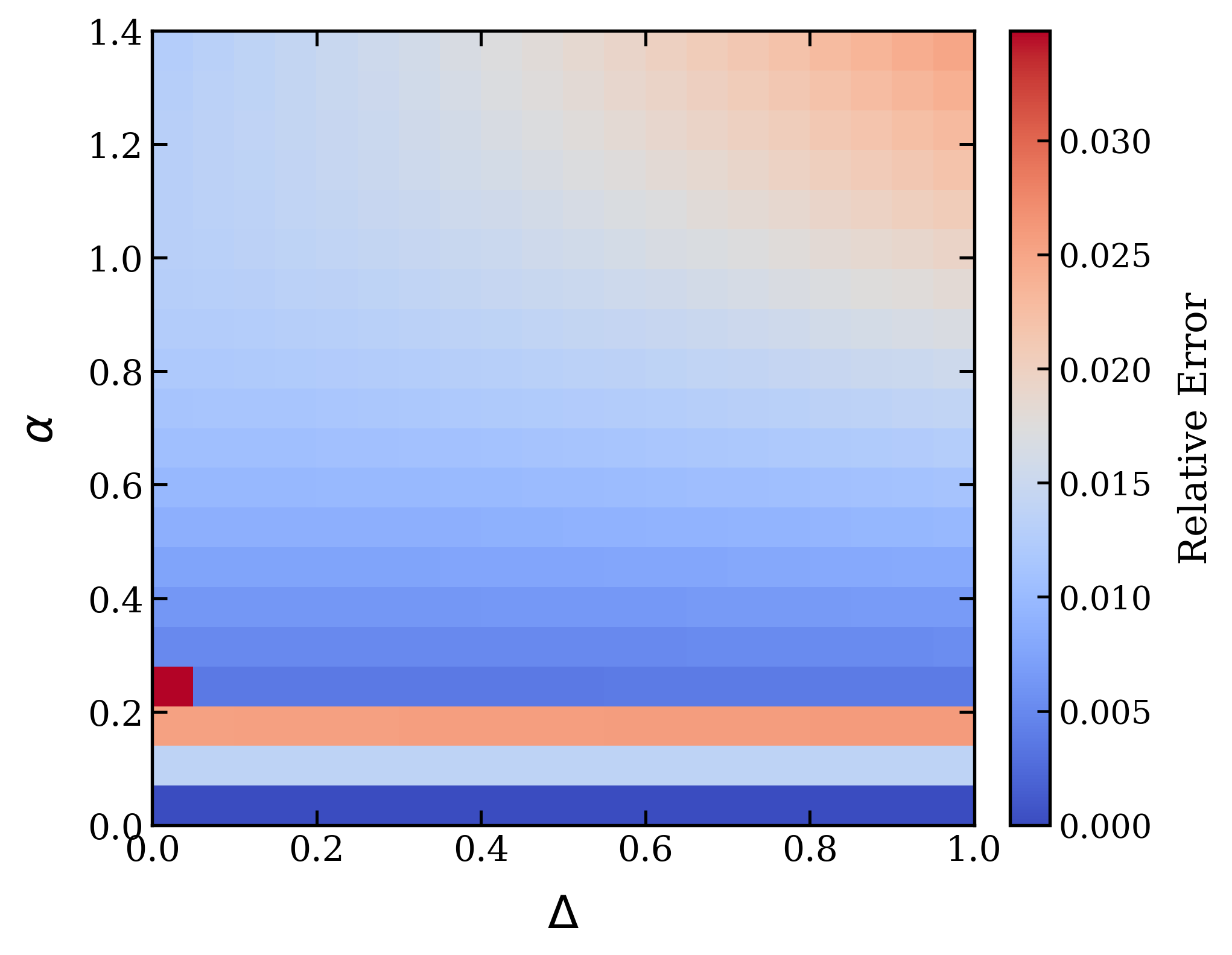}
    \caption{The plot of relative error of VQE energy with respect to the exact diagonalization in the PM phase of LRXXZ with $J=-1$.  }
    \label{fig:ferro3d}
\end{figure}
\begin{figure}[ht]
    \includegraphics[width=0.8\linewidth]{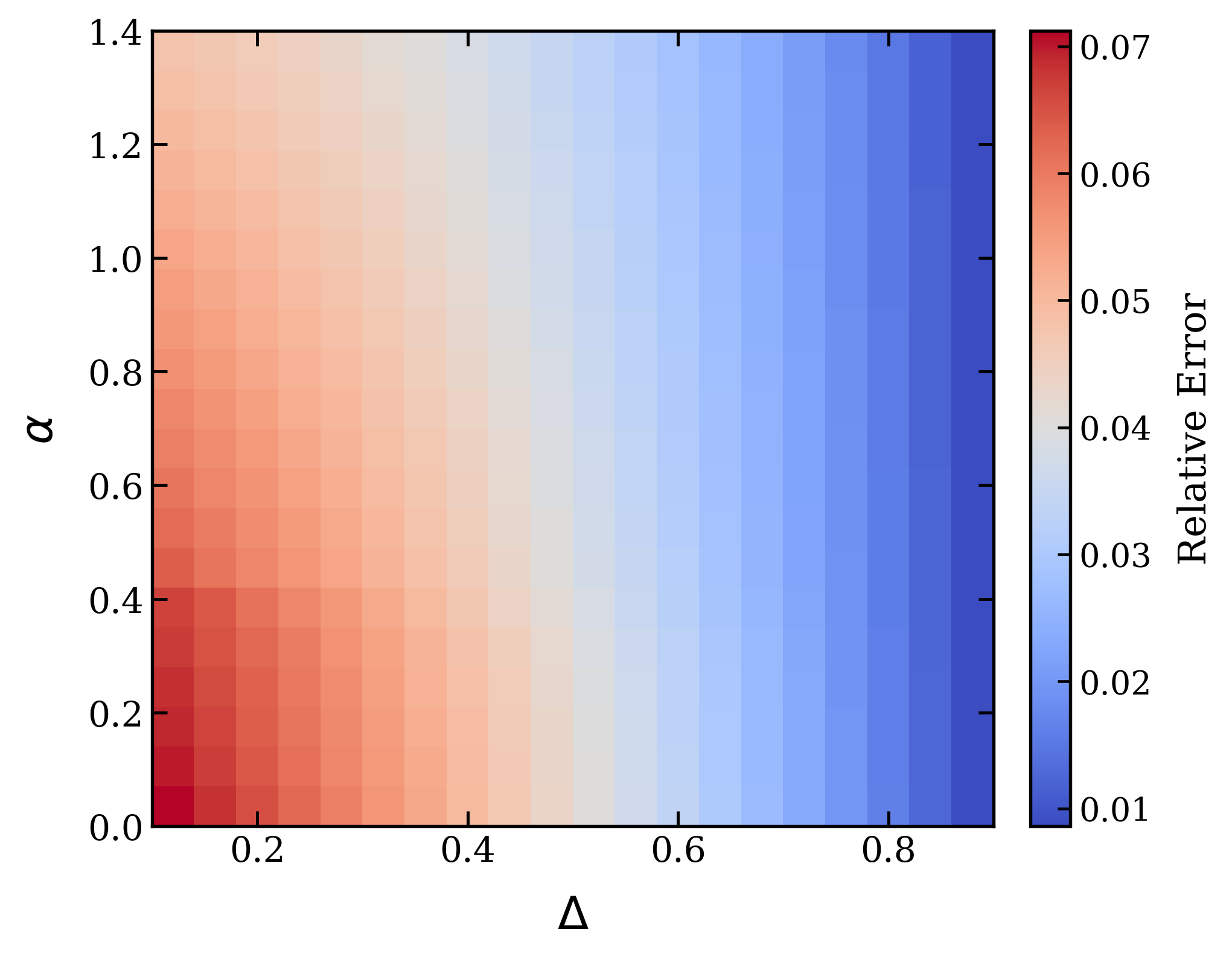}
    \caption{The plot of relative error of VQE energy with respect to the exact diagonalization in the PM phase of LRXXZ with $J=1$.  }
    \label{depth2}
\end{figure}
\subsection{Depth 2}\label{dep2}
In this section, depth 2 has been employed to accurately determine the ground state energy. In the previous section, we used the error in the VQE algorithm (with respect to the actual ground state), to study different phases of LRXXZ, but in fact we can also use the same ansatz circuit at a higher depth to find the ground state energy more accurately. In this case, we have taken $J<0$ and $J>0$. We observed that changing the sign gave us a better ground state energy when evaluated using VQE. The reason is that the Hamiltonian with the negative J has a much higher ground state energy compared to the $J>0$ case. Therefore, it was comparatively easier for the optimiser to reach the ground state for the $J<0$ case. In fact, to achieve accuracy for the positive $J$ Hamiltonian, we applied Ry gates on our initial AFM state. This allowed us to explore a larger Hilbert space and was not restricted to a certain spin sector. In turn, this allowed us to explore states with lower energy.
The ansatz circuit is the same as previously shown in Fig.~(\ref{fig:variational_circuit}). A strategy for faster convergence is used, where the parameters of the ground state energy at one point have been passed on to the next point as the initial parameters for the next, and so on. We have used the Cobyla optimiser with 4000 iterations. We observe from Fig.~(\ref{fig:ferro3d}) that the maximum relative error for $J<0$ to be around $0.03$, which is around $3\% $ and the minimum relative error is $0$. However, for $J>0$, the maximum relative error is 0.07, and the minimum is 0.01, as shown in Fig.~(\ref{depth2}).
%It was interesting to observe that with the changed sign of J, the ground state energy was much lower, and our previous ansatz was not able to reach the ground state. Therefore, we had to add an extra RY gate to the initial}-state in order to make it work. The extra RY gate allowed us to express other states with non anti-ferromagnetic terms that lowered the energy, and we were able to reach a closer ground state energy.
This reflects that our VQE works well for a wide range of parameters, even in depth 2, which has never been reported in previous studies concerning this model \cite{wiersema2020exploring,wang2023scalable}.

\section{conclusion}
In this paper, we have shown that the VQE approach is not just limited to finding the ground state energy of a system, but can be used to identify phase transitions of model Hamiltonians, which are rather difficult to identify.
We have been able to clearly identify both first-order and infinite-order phase transitions for the LRXXZ model, using the difference between the exact ground state energy and the ground state energy evaluated via VQE, referred to as the error in energy. We observe that the gradient vector of the error showed a clear distinction between different phases. Their behaviour and direction are in compliance with the mean-field approach.While noise effects have not been considered in this work and remain an important direction for future study, the present method is advantageous as it only requires knowledge of the ground state energy. Achieving an infinite-order phase transition using the ground state energy may seem counterintuitive and impossible, but through this method, it has been demonstrated to be achievable.
We should also point out that the use of directional coherence was just a mere tool to analyse the difference in the behaviour of error in both regions. One may choose any other manner in which they may identify the change of behaviour. For example, during our analysis, we had seen some signature of transition in the $E_d$ phase diagram itself, we chose to work with directional coherence.
This method can be further utilised to identify other transitions where the ground state symmetries change across the transition boundary. Depending upon the type of system and transitions, one may design their ansatz to suit the problem, and therefore, it opens up a completely new approach to identifying phase transitions.

\section{ACKNOWLEDGMENTS}
The authors thank  Min-Fong Yang for his helpful comments on the paper. Shraddha Sharma acknowledges the financial support from the Department of Science and Technology (DST), Government of India, through the DST-INSPIRE Faculty Fellowship (DST/INSPIRE/04/2023/00184; Faculty Registration No. IFA23-PH303).

\appendix

\section{System Size Dependence}\label{size}

\begin{figure}[h]
    \centering
    \includegraphics[width=0.45\textwidth]{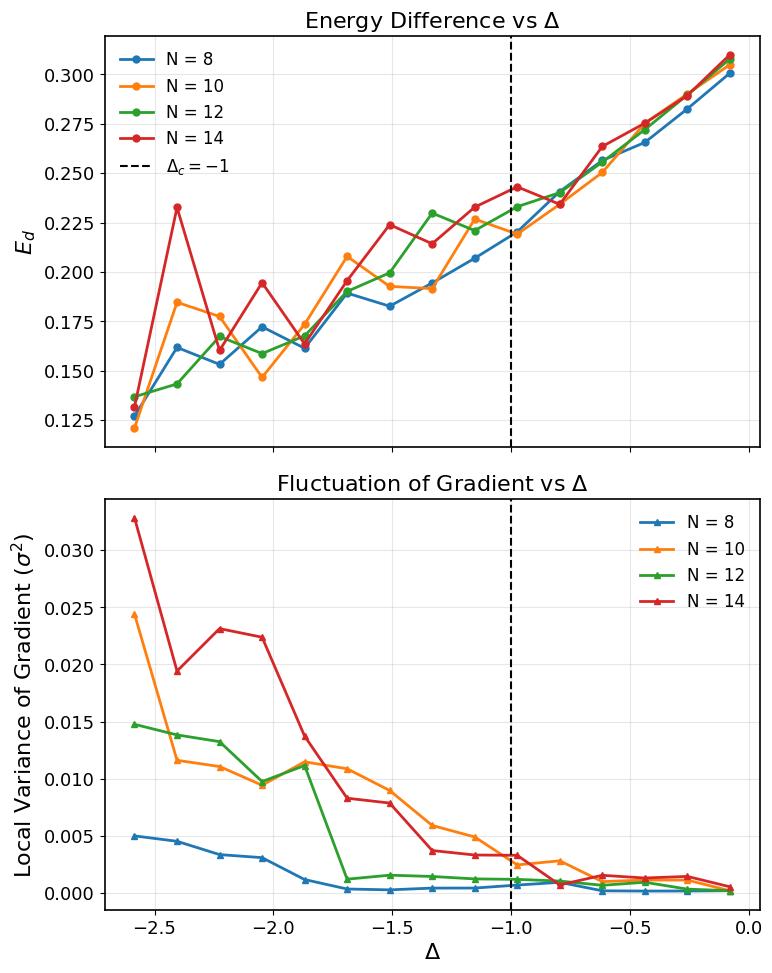}
    \caption{Top panel: energy difference versus $\Delta$ for different system sizes. Lower Panel: The plot of local variance of gradient of $E_d$ w.r.t. $\Delta$ exhibiting diminishing variance in PM phase as compared to AFM phase. The results are compared for a range of system sizes $N=8,10,12,14$ and $\alpha=4$. }
    \label{ED}
\end{figure}

The system size used for the results in the main text is $N=12$, which is sufficient for reliable VQE results\cite{wang2023scalable,de2018study}. In order to ensure that the reported phase-transition signatures and phase diagrams are reasonably converged with respect to N, in  Fig.~(\ref{ED}), we present the behaviour of $E_d$ with respect to $\Delta$, for different system sizes, $N=8,10,12$, and $14$, keeping $\alpha=4$ fixed. The range of $\Delta$ is chosen to probe the behaviour of $E_d$ across the AFM to PM phase transition occurring at $\Delta=-1$. In the upper panel of Fig.~(\ref{ED}), one can observe that $E_d$ displays visible fluctuations until $\Delta$ close to $-1$ followed by a comparatively smoother and increasing trend thereafter. This behaviour is expected since the initial state is chosen as the ground state in the AFM phase, consequently, the approximation error increases as the system enters the PM region. In the AFM region, the VQE ansatz captures the ground state reasonably well, leading to small non-systematic variations in $E_d$, whereas near and beyond the phase transition, the mismatch increases. To further quantify this, in the lower panel of Fig.~(\ref{ED}), we have used the variance of the $E_d$ gradient  over a moving window, since directional coherence cannot be computed for a single value of $\alpha$. The size of the moving window is taken to be the same as in the main text, which is equal to 10. We observe that the local variance of the gradient of $E_d$ starts to decrease from the AFM region and further saturates to zero in the PM region as the $\Delta$ is varied. 

\section{Magnitude-Weighted Directional-Coherence}\label{dir_coh}
\begin{figure}[h]
    \centering
    \includegraphics[width=0.7\linewidth]{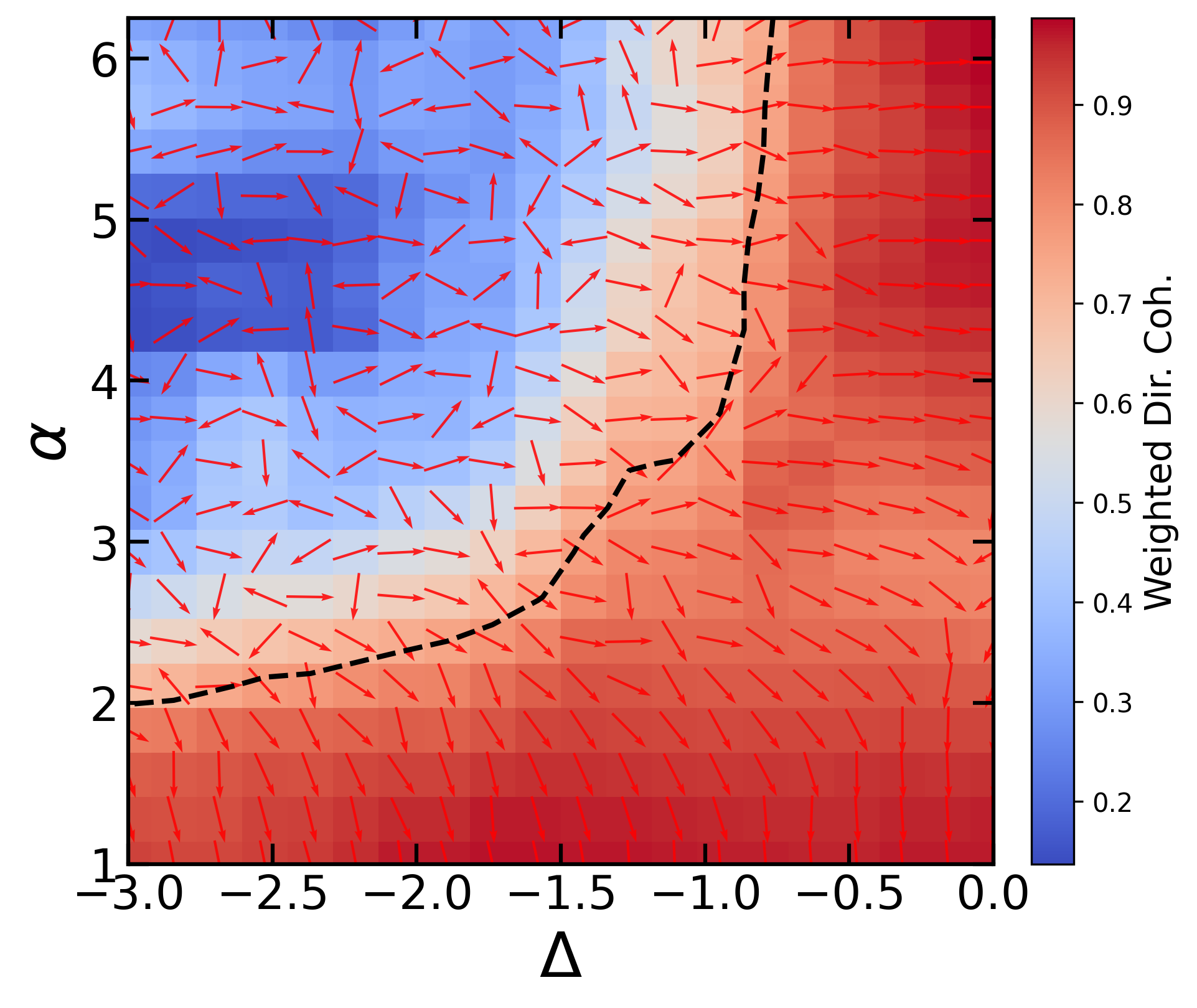}
    \caption{Magnitude-weighted directional coherence of the gradient field with parameters $(\Delta,\alpha)$ depicting AFM to PM phase transition.}
    \label{fig:dircoh}
\end{figure}

It is clear from Eq.~(\ref{uxuy}) that in the present work, we are considering the normalized components of the gradient and not really taking into account the magnitude of the gradient. Here, the gradient vector of the energy difference ($E_d(\Delta,\alpha)$) is normalized before calculating the directional coherence. As a result, the analysis intentionally retains only the direction of the gradient field. This choice was made because the objective of our method is to detect changes in the orientation pattern of the gradient field across the phase diagram, rather than the absolute scale of the variation in $E_d$. It may appear that regions with very small gradient magnitude could still exhibit high directional coherence if the vectors happen to align. However, in the present work, the phase boundaries are not determined by isolated local regions but rather by extended structure in the directional field across the ($\Delta,\alpha$) plane by taking the window size to be half of the $20\times20$ grid of ($\Delta,\alpha$). This makes the identification robust.
In order to show the robustness of this approach, we also calculate the directional coherence weighted with the magnitude given by the formula,
\begin{eqnarray}
   &~&\text{Weighted directional}\nonumber\\ 
   &~&\text{coherence}= \frac{\sqrt{(\langle|\nabla E_d| \sin{\theta}\rangle)^2+\langle|\nabla E_d| \cos{\theta}\rangle)^2}}{\langle|\nabla E_d| \rangle} \nonumber\\  \label{wDC} 
\end{eqnarray}

\noindent and show the resulting phase-diagram in Fig.~(\ref{fig:dircoh}). This result is obtained by keeping all the parameters the same as Fig.~(\ref{fig:AFM_PM}) of the manuscript (i.e., the grid size, the range of $\alpha$, $\Delta$, window size, etc.), except for using weighted directional coherence in Eq.~(\ref{wDC}) rather than Eq.~(\ref{dc}).  It is to be noted that the phase boundary between AFM-PM doesn't change significantly.  
% To support the directional coherence even further a magnitude-weighted directional coherence has also been calculated with the changed formula
% given by Eq.[\ref{DC}]
% \begin{equation}
% \text{Weighted directional coherence}= \frac{\sqrt{(\langle|\nabla E_d| \sin{\theta}\rangle)^2+\langle|\nabla E_d| \cos{\theta}\rangle)^2}}{\langle|\nabla E_d| \rangle}   \label{DC}
% \end{equation}

% The phase diagram is shown in Fig.~\ref{fig:g}. For this figure, we have kept everything the same as Fig. \ref{fig:AFM_PM}, i.e., the grid size, the range of $\Delta$, window size, etc., and just plotted the weighted directional coherence from Eq. of the manuscript for the pseudo-colour map. It is to be noted that the figure does not change much with the changed formula.

\section{Numerical Details}
For a clear understanding of readers, in this section, we provide more details on the numerical implementation specifics and parameters.
\subsection{Initial state dependence}\label{ferromagnetic}
\begin{figure}[h]
    \centering
    \includegraphics[width=0.8\linewidth]{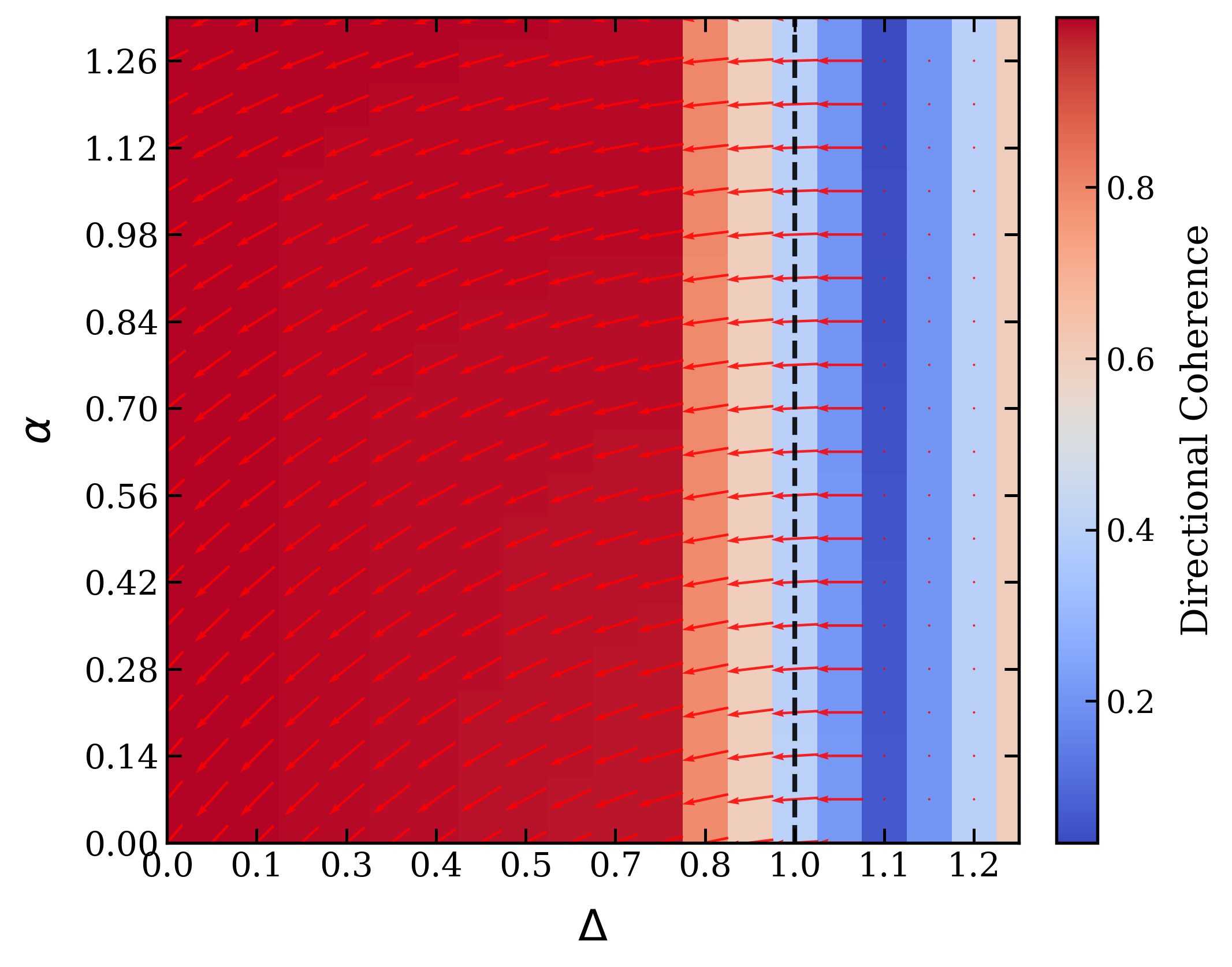}
    \caption{The PM-FM phase transition starting from FM initial state.}
    \label{fig:FM_IS}
\end{figure}

% In the manuscript, we had chosen to start with the AFM initial state, but one can choose to start with FM initial state as well. The difference will be less errors in the ferromagnetic region and more errors in the paramagnetic region as shown in the figure.
It is important to emphasize that the present technique for obtaining phase information relies heavily on the fact that the initial state for VQE should correspond to one of the two phases under consideration. Since, in that case the VQE predicted ground state energy will have less error in one phase when compared to the other phase, which will show an increase in energy difference (or error). The manuscript targeted the AFM initial state. However, to probe the PM-FM transition boundary, we can also start with the FM initial state. In Fig.~\ref{fig:FM_IS}, we present the phase diagram when we start with the FM state as an initial state. One can observe that the phase boundary doesn't change, and we obtain a valid phase transition even with the FM initial state.
In Fig.~\ref{fig:FM_IS}, the dashed vertical line is at $\Delta\approx 0.958$, which is obtained from a threshold criterion applied to the change in direction of the gradient vectors (red arrows). We apply this same procedure as we do in all our previous plots, that is, compute direction coherence over a moving window. In the figure \ref{fig:FM_IS}, one can also notice the gradient absolutely vanishing with no arrows after $Delta=1$. This is because we started with the FM initial state, and the energy calculation is very accurate in the FM region, resulting in zero error with no variation. This also shows there is no one single detector for the phase transitions, the errors may change in any specific way depending on the region, we may devise tools accordingly to analyse these changes in error.

\subsection{Implementation Specifics}\label{window}
The computation of directional coherence involves several numerical choices including the grid resolution in $\Delta$ and $\alpha$, the step sizes $\delta\Delta$ and $\delta\alpha$, the finite-difference scheme employed for gradient evaluation, and the shape, size, and weighting of the moving window, which may influence the resulting coherence maps and the inferred phase boundaries. To make this procedure fully reproducible and to establish the robustness of our results, we detail these implementation parameters here and present a sensitivity analysis with respect to the window size.
To address the sensitivity, we must observe that we utilize Eq.~(\ref{uxuy}), the components of the energy gradient vectors with respect to $\Delta$ and $\alpha$, as the red arrows in our phase diagrams (Fig.~(\ref{gradient_fm}) and (\ref{gradient_afm})). Therefore, the arrows are computed for each $\Delta$ and $\alpha$ specifically and for a specific critical point, i.e., either FM-PM transition in Fig.~(\ref{gradient_fm}) or  AFM-PM transition in Fig.~(\ref{gradient_afm}) using NumPy’s np.gradient. Therefore, the gradient vector's profile does not depend on any specific window size defined.

% \textcolor{red}{SENSITIVITY TO WINDOW SIZE PLOTS}}
\begin{figure}[h]
\includegraphics[width=0.5\linewidth]{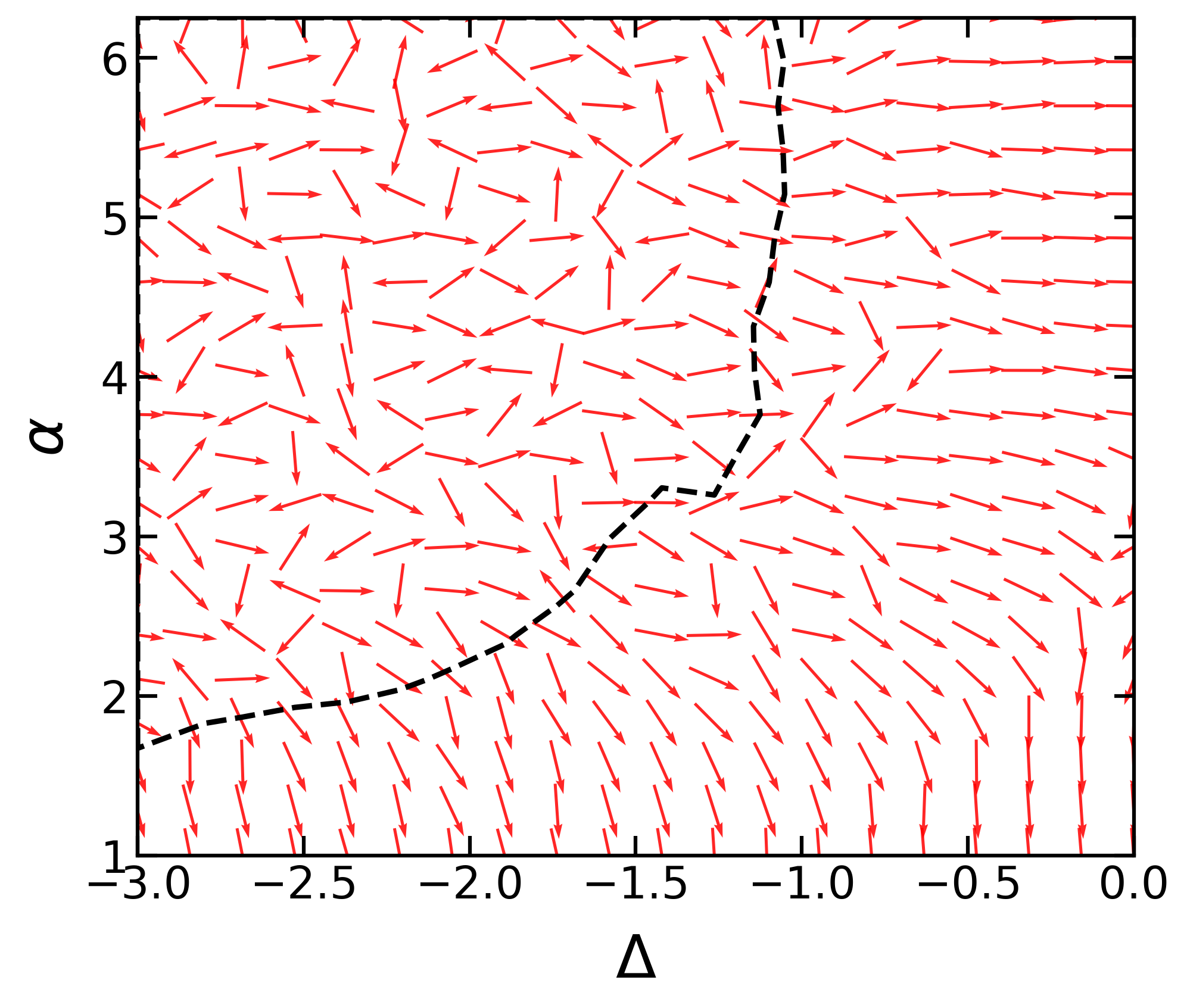}\caption{The plot of energy difference-gradient vectors calculated using NumPy’s np.gradient overlayed with a dotted contour line separating the aligned with noisy region.}
    \label{fig:vecfield}
\end{figure}

\begin{figure}[htbp]
    \centering
    \subfloat[window size$=6\times1$]{\includegraphics[width=0.48\linewidth]{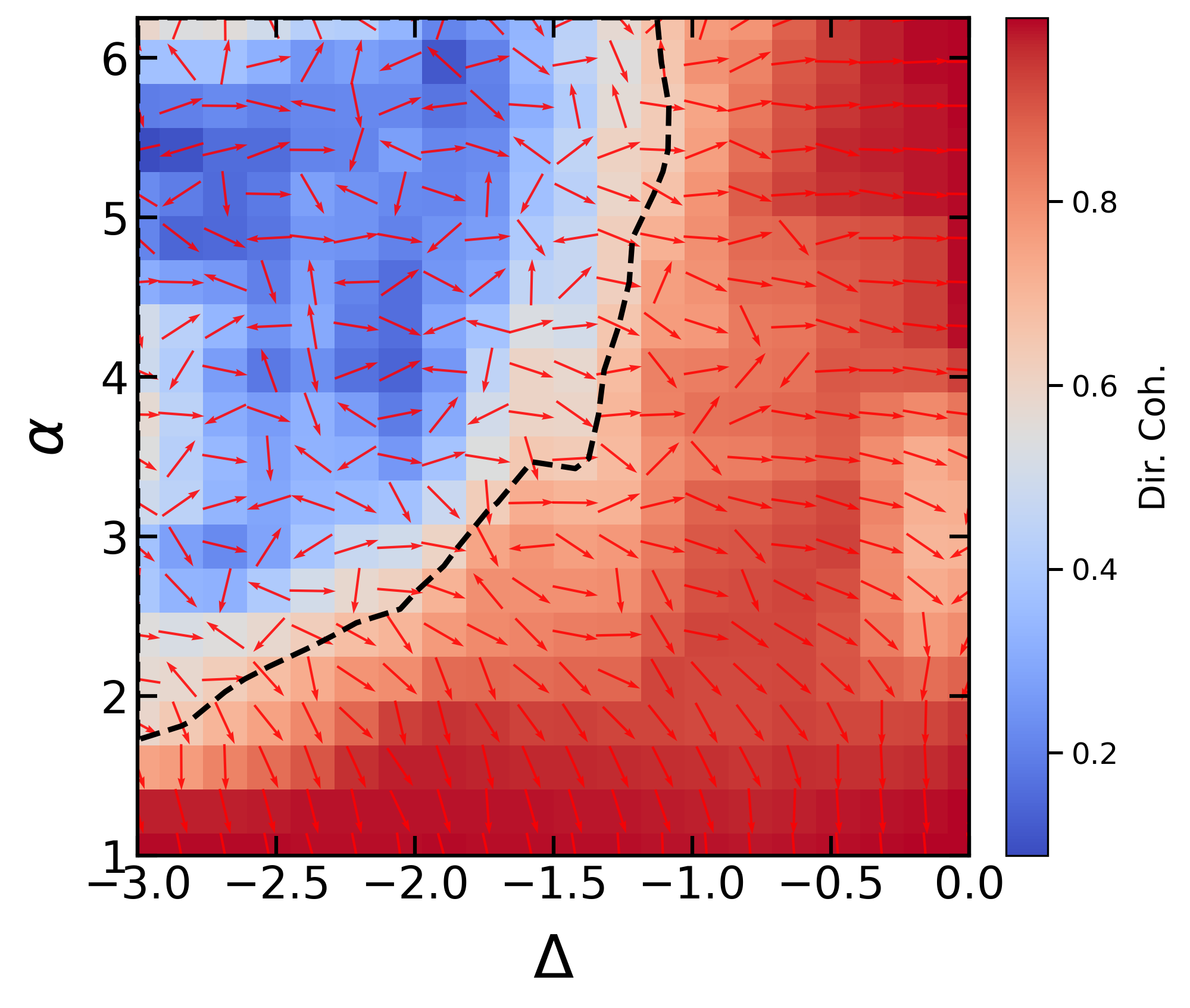}}
    \subfloat[window size$=8\times1$]{\includegraphics[width=0.48\linewidth]{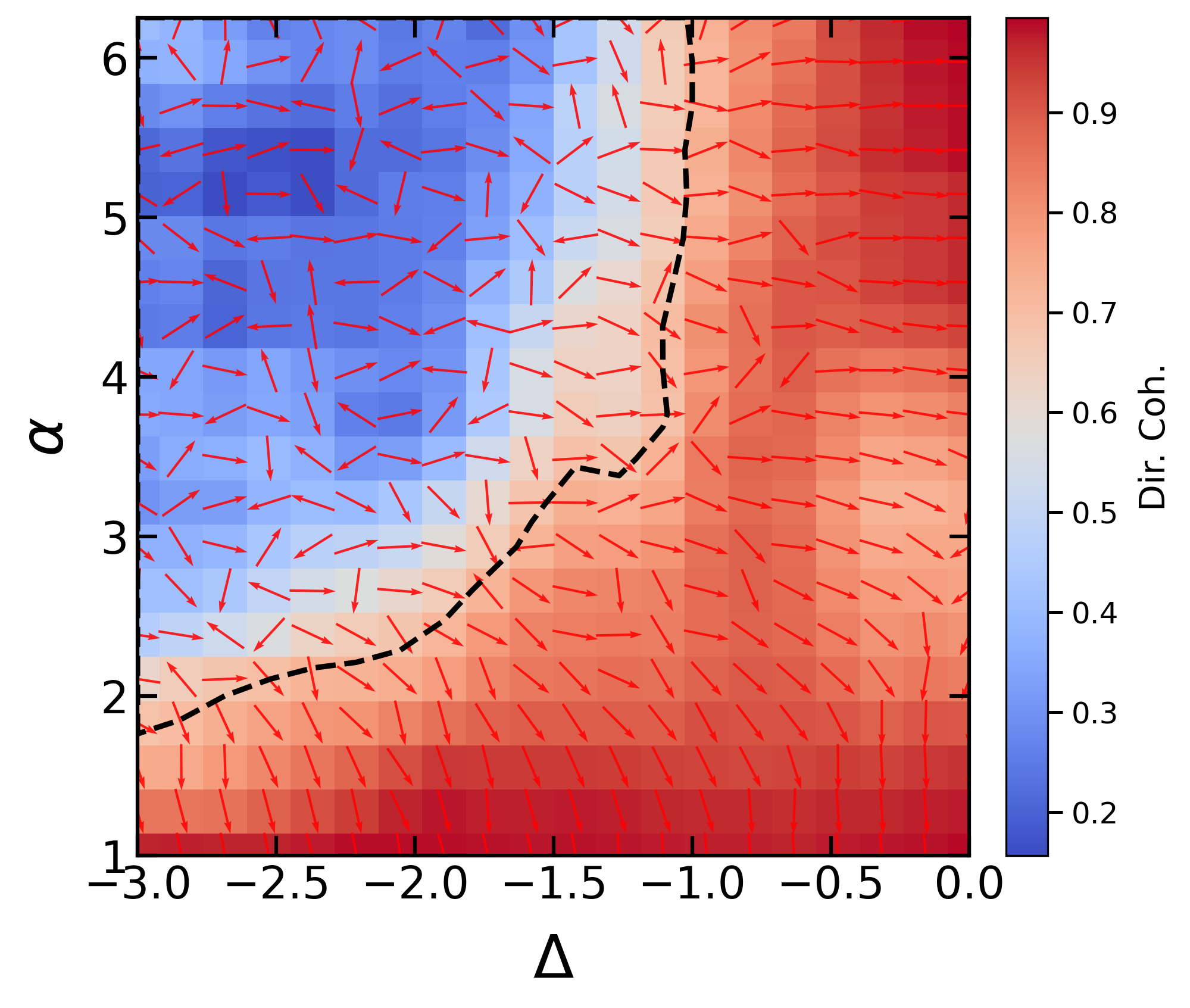}}
    \vspace{0.05em}
    \subfloat[window size$=9\times1$]{\includegraphics[width=0.48\linewidth]{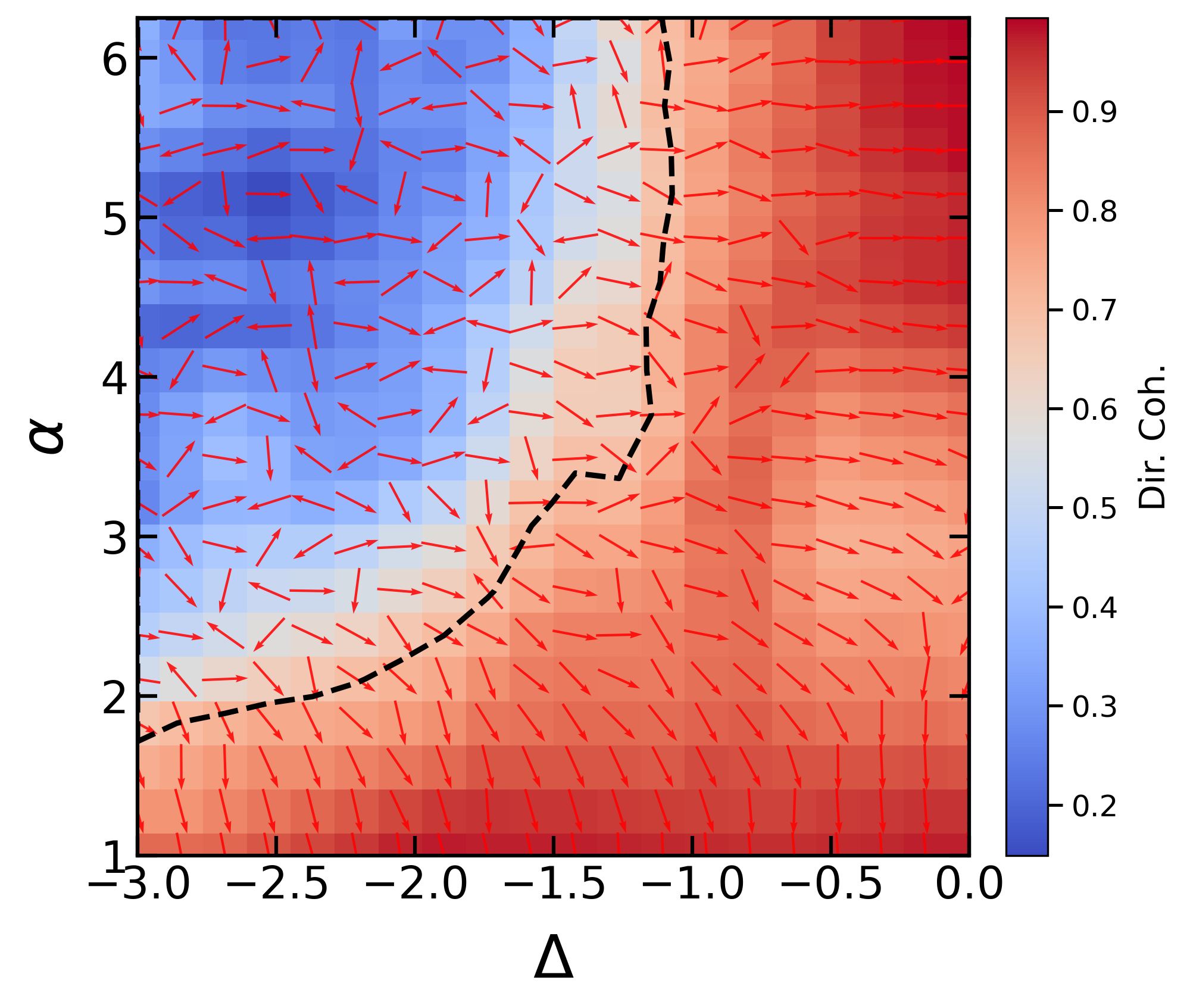}}
    \subfloat[window size$=10\times1$]{\includegraphics[width=0.48\linewidth]{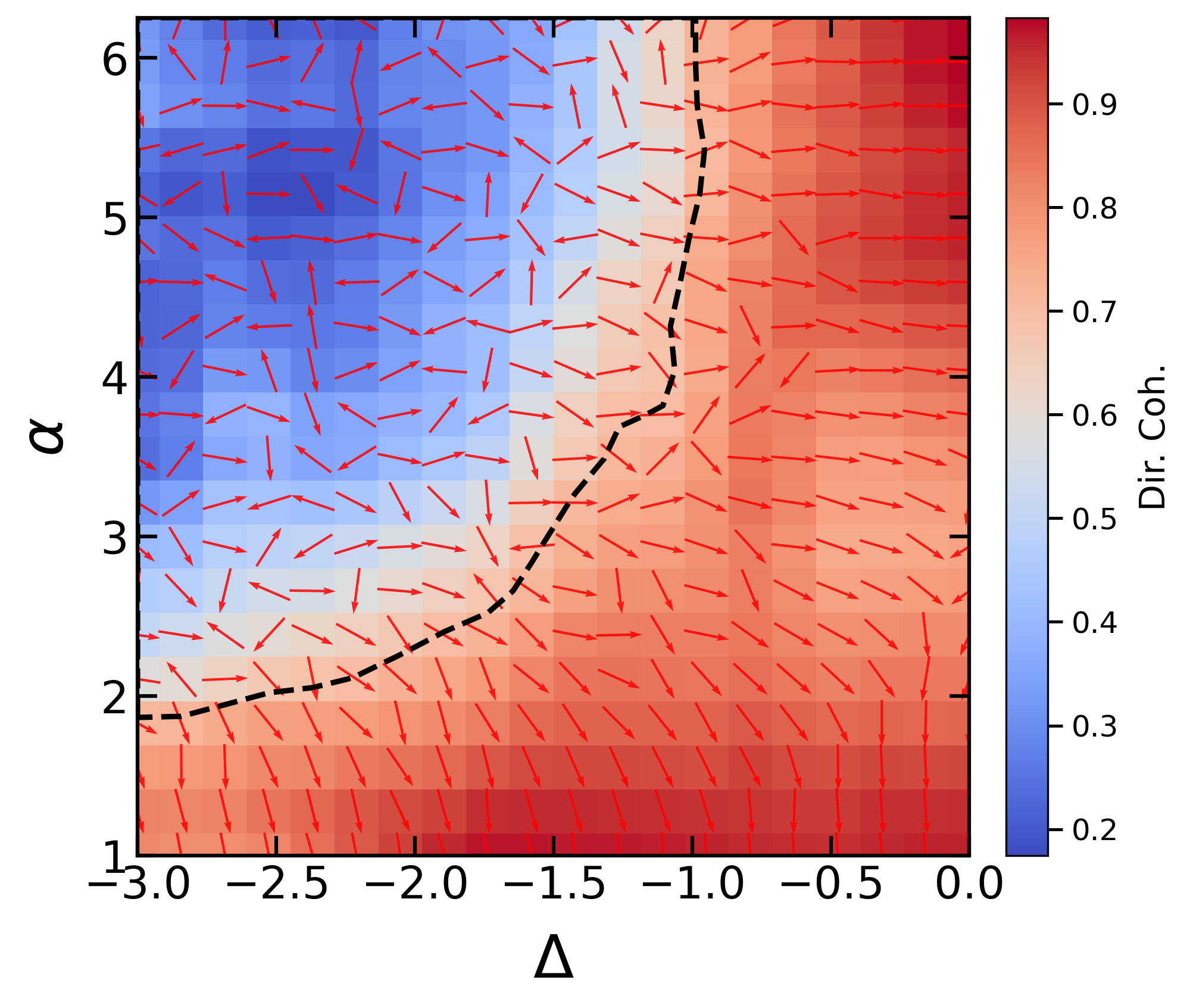}}
    \caption{The plots of AFM-PM phase transition with different window sizes.}
    \label{fig2:WS}
\end{figure}

The Eq.~(\ref{dc}), on the other hand, characterizes the pseudo-colormap of the directional coherence; naively, it separates the phases in the phase diagram where the arrows are coherent (aligned) or misaligned with respect to each other. The directional coherence, however, depends on the window size. In the present article, for a $20\times20$ grid, we have taken this window size to be half of the grid size, i.e., 10. This is solely done to clearly separate the noisy (misaligned) region and the coherent region. Once the gradients (arrows-based diagrams) are obtained, one can manually adjust the window size for a clear graphical representation. What matters is that this should align well with the overlaid vector field (arrow's direction). Therefore, the underlying principle of utilizing errors as a probe does not depend on or change with any window size, and the directional coherence is just a tool for visualization and clearly separating the two regions.
As shown in Fig.~(\ref{fig:vecfield}), the vector field (arrows, obtained from the energy gradient) shows an alignment in the PM region and a noisy (misaligned) behavior in the AFM region. The dotted-contour line is only a tool that clearly marks the separation between the two regions. The main idea of this work is to highlight that even though there exists no known way to utilise ground-state energy in order to probe an infinite-order phase transition, this technique of utilizing errors by specifically choosing an initial state can mark the phase boundary. In Fig.~(\ref{fig2:WS}), we present results for different window sizes varying from $6\times1$ to $10\times1$ and show that the deviation in the phase boundaries as compared to the $10\times1$ case (used in this work) is small.
To specify the details, the ranges of $\Delta$ and $\alpha$ are taken such that one only targets one phase transition boundary, as depicted in Fig.~(\ref{fig:FM_PM})   and Fig.~(\ref{fig:AFM_PM}). These $\alpha$ and $\delta$ values are taken in a grid of $20\times20$, the step sizes for Fig.~(\ref{fig:FM_PM}) is, $\delta\Delta=1.3/20=0.065$ and $\delta\alpha=(1.4-0.01)/20\sim0.07$.  for Fig.~(\ref{fig:AFM_PM}) of the manuscript is $\delta\Delta=3/20=0.15$ and $\delta\alpha=(5.5-1)/20\sim0.225$. The finite-difference scheme used for the gradients is NUMPY's np.gradient, and the moving-window parameter's size is $10\times1$, with no weights, and no additional smoothing/interpolation is used in these plots.
\bibliographystyle{apsrev4-2}
\bibliography{cite}
\end{document}